\documentclass[aps,pre,twocolumn,showpacs,preprintnumbers,amsmath,amssymb,superscriptaddress]{revtex4}
\usepackage{graphicx}
\usepackage{bm}
\usepackage{psfrag}
\usepackage{subfigure}
\usepackage{color}
\usepackage[normalem]{ulem}

\newcommand{\be}{\begin{equation}}
\newcommand{\ee}{\end{equation}}
\newcommand{\ba}{\begin{eqnarray}}
\newcommand{\ea}{\end{eqnarray}}

\newcommand{\phiC}{\phi_{\scriptscriptstyle\rm C}}
\newcommand{\phiJ}{\phi_{\scriptscriptstyle\rm J}}

\newcommand{\ma}[1] {\textcolor{black}{#1}}
\newcommand{\maa}[1] {\textcolor{black}{#1}}

\newcommand{\ket}[1]{|#1\rangle}
\newcommand{\bra}[1]{\langle #1|}

\newcommand{\sFrac}[2]{{\textstyle\frac{#1}{#2}}}
\renewcommand{\th}{^{\mbox{\tiny th}}}

\begin{document}
 \author{Carolina Brito}
\affiliation{Instituto de Física, UFRGS, 91501-970, Porto Alegre, Brazil}
 \author{Edan Lerner}
 \affiliation{Institute for Theoretical Physics, University of Amsterdam,
Science Park 904, 1098 XH Amsterdam, The Netherlands}
\author{Matthieu Wyart}
\affiliation{Institute of Physics, EPFL, CH-1015 Lausanne, Switzerland}

\date{\today}

\title{Theory for Swap Acceleration near the Glass and Jamming transitions}

\begin{abstract}
  Swap algorithms can shift the glass transition to lower temperatures, a recent unexplained observation  constraining the nature of this phenomenon. Here we show that swap dynamic is governed by an effective potential describing both particle interactions as well as  their ability to change size. Requiring its stability is more demanding than for the potential energy alone. This result  implies that stable configurations appear at lower energies with swap dynamics,  and thus at lower temperatures when the liquid is cooled.   \maa{ The magnitude of this effect is proportional to the width of the radii distribution, and  decreases  with compression for finite-range purely repulsive interaction potentials.} We test these predictions numerically and discuss the implications of these findings for the glass transition.We extend these results to the case of hard spheres where swap is argued to destroy meta-stable states of the free energy coarse-grained on vibrational time scales. Our analysis  unravels the soft elastic modes responsible for the speed up swap induces, and allows us to predict the structure and the vibrational properties of glass configurations reachable with swap. In particular for continuously poly-disperse systems we predict the jamming transition to be dramatically altered, as we confirm numerically.  A surprising practical outcome of our analysis is new algorithm that generates ultra-stable glasses  by simple descent in an appropriate effective potential. 

\end{abstract}


\maketitle

\section{Introduction}

Understanding the mechanisms underlying the slowing down of the dynamics near the glass transition is a long-standing challenge in condensed-matter \cite{Ediger96,Angell91}. Unexpectedly, swap algorithms \cite{Grigera01c, Fernandez06} (in which particles of different radii can swap in addition to the usual  moves of particle positions) were recently shown to allow for equilibration of liquids far below the glass transition temperature $T_g$  \cite{Berthier16b,Gutierrez15,Ninarello17,Berthier17}.   For judicious choice of poly-dispersity, one finds that: $(i)$ the glass transition is shifted to lower temperatures: with swaps the $\alpha$-relaxation time at $T_g$ is only two or three order of magnitudes slower that in the liquid, instead of 15 orders of magnitude for regular dynamics. The slowing down of the dynamics occurs at a lower temperature, which we refer to as $T_0^{\mbox{\scriptsize swap}}$. $(ii)$~The spatial extent of dynamical correlations, which are significant near $T_g$, are greatly reduced with swap and only occur at $T_0^{\mbox{\scriptsize swap}}$.  $(iii)$ The mean square displacement  of the particles on vibrational time scales is increased significantly in this temperature range \cite{Ninarello17}.  These observations constrain theories of the glass transition. In particular, current formulation of theories based on a growing thermodynamic length scale appear inconsistent with these observations \cite{Wyart17}. 
A theory of the glass transition should explain both swap and non-swap dynamics. % Although propositions for the latter abound, no mechanisms have been yet proposed to explain the speed up of dynamics induced by swap moves.
 Goldstein \cite{Goldstein69} proposed that the glass transition is initiated by a transition in the free energy landscape: at high temperature, the system resides near saddles, whereas below some temperature $T_0$ the dynamics can only occur by activation (whose nature is  debated), and is thus much slower.  In mean-field models of structural glasses such a transition in the landscape is predicted \cite{Lubchenko07,Broderix00,Grigera02, Kurchan16} and corresponds to a Mode Coupling Transition (MCT) where the relaxation time diverges. 
 It was suggested that the MCT transition would be shifted to lower temperature with swap dynamics in  \cite{Wyart17}, as proven and confirmed numerically in a mean-field model of glasses \cite{Ikeda17}. Yet, understanding the real-space mechanisms underlying the speed up induced by swap in finite dimensions (where the relaxation time cannot diverge) as well as the nature of the very stable glassy configurations  swap can reach remains a challenge.

In this work, we tackle these questions by first reviewing the equilibrium statistical mechanics theory of polydisperse systems \cite{Glandt_1984,Glandt_1987}, to show that they are equivalent to a system of identical particles that can individually deform according to a chemical potential $\mu(R)$, where $R$ is the particle radius. %This equivalence applies in the thermodynamic limit, where $\mu(R)$ is a chemical potential ensuring that the correct poly-dispersity $\rho(R)$  is obtained. 
In the (practically important) case where poly-dispersity is continuous, $\mu(R)$ is smooth, allowing us to define  normal modes of the generalised Hessian that includes radii as degrees of freedom.  We prove that requiring its stability is strictly more demanding than for the usual Hessian. Second, we show that these results stringently constrain the glassy states generated by swap algorithms. We illustrate this point by studying the jamming transition in soft repulsive particles, which we prove must be profoundly altered: hyper-staticity  is found with an excess number of contacts  $\delta z$ with respect of the Maxwell bound  $\delta z\!\sim\!\alpha^{1/2}\!>\!0$, where $\alpha$ characterises the width of the radii distribution $\rho(R)$.  Although we find that
 the vibrational spectrum of the generalised Hessian is marginally stable with respect to soft extended modes near jamming,  these modes are gapped in the regular Hessian, unlike for packings obtained with regular dynamics  \cite{Wyart05a, DeGiuli14b, Franz15}. These results are verified numerically by introducing a novel algorithm performing a steepest descent in the generalised potential energy that includes  $\mu(R)$, which can generate extremely stable glasses {\it without any activation}. Third, we investigate the glass transition.  We show that  the inherent structures obtained after a rapid quench with the regular dynamics are unstable with respect to this new algorithm, which reaches significantly smaller energies. This result indicates that metastable states appear at lower energies with swap, and therefore at lower temperatures when the liquid is equilibrated. Thus the Goldstein transition must be shifted to a lower temperature with swap dynamics, suggesting a natural explanation for its speed up which specifies the collective modes  facilitating the dynamics for $T_0^{\mbox{\scriptsize swap}}<T<T_g$. \maa{ We predict this shift  to be proportional to $\alpha$ in general, and  to be inversely proportional to the distance to jamming for sufficiently compressed soft spheres.} Lastly, we argue that these results apply to hard spheres as well, if the energy is replaced by a coarse-grained free energy landscape as previously studied in \cite{Brito06,Brito09,DeGiuli14b,Altieri16}. We use this approach to provide a simple phase diagram  where  the Goldstein transition and the emergence of marginality \cite{Wyart05a} (referred to as  a Gardner transition in infinite dimension \cite{Altieri16}) can be related to structure for both swap and non-swap dynamics.

\section{Grand-canonical description of poly-disperse systems}
\subsection{Effective potential}

We now show  that a poly-disperse system can be described by an effective potential that includes the radius as degree of freedom, an idea already present in the early works of \cite{Glandt_1984,Glandt_1987}.  We consider a system of $N$ particles with continuous polydispersity $\rho(R)$, of width $\alpha\!=\!\langle (\langle R^2 \rangle\!-\!\langle R \rangle ^2)^{1/2}\rangle/\langle R\rangle$. Here $\{R\}$ indicates the set of particle  radii and $\{r\}$ their positions. In what follows we denote $\langle R\rangle \equiv R_0$, and ${\cal U}(\{r\},\{R\})$ the total potential energy in the system. We define the partition function $Z(\{r\})$ annealed over the particle radii:
\be
\label{1}
Z(\{r\})=\sum_{\cal P(\{R\})} \exp\big(-\beta  {\cal U}(\{r\},\{R\})\big)\,,
\ee
where the sum is on all the permutations ${\cal P}(\{R\})$ of the particle radii.  In the thermodynamic limit, a grand-canonical formulation is equivalent, in which particles of different radii correspond to different species. The associated partition function writes:
\be
\label{2}
Z_{GC}(\{r\})=\sum_{\cal \{R\}} \exp\bigg[-\beta  \bigg({\cal U}(\{r\},\{R\})+\sum_i^N \mu(R_i)\bigg)\bigg] \,,
\ee
where $\mu(R)$ is the chemical potential at radius $R$. It is chosen such that in the thermodynamic limit, 
the distribution of radii  that follows from Eq.~(\ref{2}) is 
\begin{eqnarray}
\rho(R) &\equiv &\frac{1}{Z} \sum_{{ \{r\}}, \cal \{R\}} \frac{1}{N}\bigg(\sum_i \delta(R-R_i)\bigg) \nonumber \\
& & \times\exp\bigg[-\beta  \bigg({\cal U}(\{r\},\{R\})+\sum_i^N \mu(R_i)\bigg)\bigg].
\end{eqnarray}
%\sout{$\rho(R)$ \footnote[4]{ $\rho(R)\equiv\frac{1}{Z} \sum_{{ \{r\}}, \cal \{R\}} \frac{1}{N}[\sum_i \delta(R-R_i)]  \exp[-\beta  ({\cal U}(\{r\},\{R\}) +\sum_i^N \mu(R_i))]$.}}
%A priori $\mu(R)$ depends on the particle positon $\{r\}$. However when considering the motion of a finite subset of particles (as relevant for stability),  $\mu(R)$ is fixed since it is an intensive quantity. 
A key remark is that once Eq.~(\ref{2}) is integrated on particle positions $\{r\}$, one obtains the partition function  for the coupled degrees of freedom $\{r\}$ and $\{ R\}$ with an effective energy functional:
 \be
 \label{5}
 {\cal V}(\{r\},\{R\})={\cal U}(\{r\},\{R\})+\sum_i^N \mu(R_i).
 \ee

\subsection{Mechanical stability under swap}
Let us consider first an athermal system. In the thermodynamic limit, mechanical stability under swap dynamics  requires ${\cal V}$ to be at a minimum. Beyond the usual force balance condition, it implies:

\be
\label{4}
\frac{\partial {\cal U}}{\partial R_i}\equiv \sum_j f_{ij}=-\frac{\partial \mu}{\partial R}\bigg|_{R=R^*_i}
\ee
where $f_{ij}$ are the contact forces between particle $i$ and $j$ (positive  in our notations for repulsive forces), and ($\{r^*\}$, $\{R^*\}$) the particle positions and radii at the minimum. For unimodal distribution $\rho(R)$, one expects $\mu(R)$ to be unimodal too. In an amorphous solid the fluctuations of the  left hand side of Eq.~(\ref{4}) are of order  $p R_0^{d-1}$ where $p$ is the pressure and $d$ the spatial dimension.  To achieve a distribution of radii of width $\alpha$, the stiffness $k_R$ acting on each particle radius must thus be of order:
\be
\label{111}
k_R\!\equiv\! \langle \partial^2 \mu/\partial^2 R \rangle_i \!\sim\! p R_0^{d-2} /\alpha,
 \ee
   where the average is made on all particles i.

\subsection{Generalised vibrational modes}
Stability also requires the Hessian $H_{\mbox{\scriptsize swap}}$ (the matrix of second derivatives of $ {\cal V}$) to be positive definite. Since they are now $N(d+1)$ degrees of freedom, $H_{\mbox{\scriptsize swap}}$ is  a $N (d+1)\times N (d+1)$ symmetric matrix, of eigenvalues $\omega_{\mbox{\scriptsize swap}}^2$. It contains a block of size $Nd\!\times\!Nd$ which is the regular Hessian $H_{ij}\!=\!\partial^2 U/\partial r_i\partial r_j$. We denote by $\omega^2$ its eigenvalues.  Because hybridisation with additional degrees of freedom can only lower the minimal eigenvalues of the Hessian, $H_{\mbox{\scriptsize swap}}$ has lower eigenvalues than $H$ (as quantified below), implying that mechanical stability is more stringent with swap dynamics. %Thus the marginal stability line for the swap must be strictly contained in the  stable phase of non-swap dynamics, as illustrated in Fig.\ref{f1} left. 

Let us illustrate this result perturbatively when $k_R\!\gg\! k$, where $k$ is the characteristic stiffness of the interaction potential ${\cal U}$. In general, the eigenvalues of $H$ are functions of the set of stiffnesses $\{k_{ij}\}$, but also of the interaction forces $\{ f_{ij}\}$ \cite{Alexander98}. We first ignore the effects induced by such pre-stress. Moving along a normal mode of $H$ by a distance $x$ (while leaving the radii fixed) leads to an elastic energy $\sim\!\omega^2 x^2$ and change forces by a characteristic amount $\delta f$ satisfying $\delta f^2/k\!\sim\! x^2 \omega^2$. Because of such change, Eq.~(\ref{4}) is not satisfied anymore. Thus the potential ${\cal V}$ can be reduced further  by an amount of order $\delta f^2/k_R\!\sim\!\omega^2  x^2 k/k_R$ if the radii are allowed to adapt. This reduced energy  can be approximatively written as $x^2\omega_{\mbox{\scriptsize swap}}^2$, where  $\omega\mbox{\scriptsize swap}^2$ is the eigenvalue associated to that mode in the effective Hessian. We thus obtain:
\be
\label{6}
\omega^2-\omega_{\mbox{\scriptsize swap}}^2\sim \omega^2\frac{k}{k_R} \ \  \hbox{  for } k_R \gg k %\hbox{(i.e. }  \frac{pR_0^{d-2}}{k}\gg \alpha).  %sim %\omega^2 \alpha\frac{k}{p R_0^{d-2}}\\ \  \hbox{  for } k_R \gg k. %sim %\omega^2 \alpha\frac{k}{p R_0^{d-2}} 
\ee

\section{Soft sphere systems}
To illustrate these ideas, we consider soft spheres with half-sided harmonic interactions, so that:
\be
\label{pot}
{\cal U}(\{r\},\{R\})=\sum_{i,j} \frac{k}{2} (r_{ij} -R_i-R_j)^2\Theta(R_i+R_j-r_{ij})
\ee
where $r_{ij}$ is the distance between particles $i$ and $j$, and $\Theta(x)$ is the Heaviside step function. 

\subsection{Jamming transition for soft spheres under swap}
When materials with such finite-range interactions are quenched to zero temperature, they can jam into a solid or not depending on their packing fraction.  At the jamming transition separating these two regimes, vibrational properties are singular \cite{Liu10,Wyart05b}, and the effects of swap are expected to be important, as we now show.  The vibrational spectrum of the regular Hessian is strongly affected by excess number of constraints $\delta z$ with respect to the Maxwell threshold where \ma{the numbers of degrees of freedom and constraints match}. Effective medium \cite{Wyart10a} or a variational argument \cite{Yan16} imply that in the absence of pre-stress, soft normal modes  in the Hessian  must be present with eigenvalues:
\be
\label{wstar}
\omega^*{}^2\!\sim\! k ~\delta z^2.
\ee
For swap with a small poly-dispersity  $\alpha<<\Delta$, where we introduced  the dimensionless particle overlap $\Delta\equiv  pR_0^{d-2}/k$, then from Eq.\ref{111} $k_R>>k$ and  Eqs.~(\ref{6}) applies. It implies  that soft normal modes will be present at lower eigenvalues $\omega_{\mbox{\scriptsize swap}}^*{}\!^2\!\sim\! \delta z^2 k(1-C_0 \alpha/\Delta)$ where $C_0$ is a numerical constant. Pre-stress can be shown to shift eigenvalues of the Hessian by some amount $\approx \!-C_1 k \Delta$ \cite{Wyart05a,DeGiuli14}, leading to  eigenvalues satisfying  $\omega_{\mbox{\scriptsize swap}}^0{}\!^2\!\sim\!\delta z^2 k(1-C_0 \alpha/\Delta)-C_1k \Delta$. 
Mechanical stability requires positive eigenvalues  and we obtain:
\be
\label{8}
\delta z\geq \sqrt{\frac{C_1 \Delta }{(1-C_0 \frac{\alpha }{\Delta})}} \ \  \hbox{  for } \alpha<<\Delta
\ee
\maa{Eq.\ref{8} indicates that away from jamming, the relative effects of swap on the structure are proportional to $\alpha/\Delta$.}  Certain materials are marginal stable,  corresponds the saturation of inequalities of the kind of Eq.~(\ref{8}). As we shall see below, we provide numerical evidence that it is also the situation if swap is used, at least near jamming. Here this assumption gives an expression for $\delta z$ which is above (but very close to in the limit   $\alpha<<\Delta$) the bound for non-swap dynamics of \cite{Wyart05a}, recovered by setting  $\alpha\!=\!0$. Thus in this limit we expect very small change of structure in the glass phase between swap and non-swap dynamics. 

For swap with a large poly-dispersity  $\alpha>>\Delta$, the situation is completely different. 
We then have $k_R<<k$: in this regime the strong interactions correspond to interactions between particles in contact.
As far as the low-frequency end of the spectrum is concerned, these interactions can be considered to be hard constraints (i.e. $k=\infty$), whose number is $Nz/2$.
The dimension of the vector space satisfying such hard constraints is $N(d+1)\!-\! Nz/2\!=\! N\!(1-\! \delta z/2) $. These modes gain a finite frequency due to the presence of the weaker interactions of strength $k_R$ associated with the change of radius, of strength $k_R$. Importantly, the number of these weaker constraints left is simply the number of particles $N$.
If $\delta z$ is small  the number of degrees of freedom $N(1-\delta z/2)$ is just below the number of constraints $N$: for this vector space we are close to the "isostatic" or Maxwell condition where  the number of constraints and degrees of freedom match. Thus we can use the same results for the spectrum valid near the jamming transition introduced above. They also apply in that situation, with the only difference that the stiffness scale $k$ is replaced by $k_R$. In particular if pre-stress is not accounted for,
a plateau of soft modes must appear above some frequency given by Eq.(\ref{wstar}):
\be
\label{9}
\omega^*_{\mbox{\scriptsize swap}}{}^2\sim k_R~\delta z ^2 \sim k \frac{\Delta}{\alpha} \delta z^2,
\ee
 This plateau survives up to the characteristic frequency $\omega_i\!\sim\!\sqrt{k_R}$. When pre-stress is accounted for, eigenvalues of the Hessian are again shifted by $\sim\!-k\Delta$. Mechanical stability then implies $\Delta/\alpha \delta z^2\! >\! C_2 \Delta$ and:
\be
\label{10}
\delta z\geq C_2 \sqrt{\alpha} \ \  \hbox{   for } \Delta \ll \alpha
\ee
In this regime, marginal stability (the saturation of the stability bound of Eq.~\ref{10}) corresponds to a coordination independent of pressure, with $\delta z\sim \sqrt{\alpha}$ and $\omega^*_{\mbox{\scriptsize swap}}{}^2\!\sim\! k\Delta$. We thus predict that swap dynamics
destroys isostacity, and significantly affect structure and vibrations. For sufficiently large $\alpha$, this regime will include the entire glass phase, and vibrational properties and stability will be affected in the vicinity of the glass transition (which sits at a finite distance from the jamming transition \cite{Ikeda12}) as well. %This is the situation sketched in Fig.~\ref{f1}.

Note that these predictions apply to algorithms that allow for swap moves up to the jamming threshold. This is not the case e.g. in \cite{Ozawa17}, where swaps are used to generate dense equilibrated liquids, that are then quenched without swap  toward  jamming. 
We also expect  isostaticity to  be restored in algorithms for which the set of particle radii is strictly fixed, but only below some pressure $p_N$ that vanishes as $N\!\rightarrow\!\infty$, above which our predictions should apply.

%{\bf Hard spheres under swap:} The reasoning is identical to soft spheres, except that for hard spheres $(i)$ the pressure $\tilde p$ diverges as jamming is approached, and the stiffness of particle interactions diverges as $k\!\sim\!\tilde p^2$ \cite{Brito09} $(ii)$ marginal stability for non-swap dynamics corresponds to $\delta z\!\sim\!\tilde p^{-(2+2\theta)/(6+2\theta)}$ where $\theta\!\approx\!0.41$ \cite{DeGiuli15}.  Equating the interaction stiffness with $k_R$ leads to a characteristic pressure $\tilde p_s\!\sim\! 1/\alpha$ beyond which swap starts to affect structure. For larger pressure stability implies  $\delta z\!\geq\!\alpha^{(2+2\theta)/(6+2\theta)}$, again indicating that isostaticity is lost with swap.

\begin{figure}[htbp]
\centering
\includegraphics[width=1.0\columnwidth]{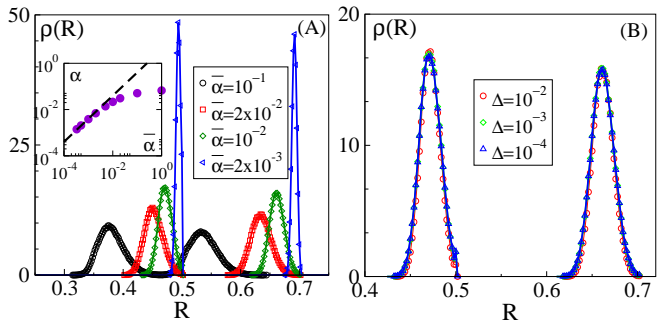}
\caption{(A) Distribution of radii $\rho(R)$ for different ${\bar \alpha}$ at fixed $\Delta=10^{-4}$ and  for different $p$ at fixed  ${\bar \alpha}=10^{-2}$. (B). Inset: $\alpha$ {\it v.s.}~${\bar \alpha}$ compared to a linear relation in black.  Data are averaged over 1000 systems with $N=484$ particles. For ${\bar \alpha}=10^{-2}$, the width of the distributions are given by $\alpha_1=0.024$ and $\alpha_2=0.018$ for the smaller and bigger radii respectively.}
%$\alpha_1=0.0114$ and $\alpha_2=0.0124$ -- these are the values before diving by R1 and R2
\label{f1n}
\end{figure}

\subsection{Numerical Model}
As shown in Eq.~(\ref{5}), swap dynamics \ma{is equivalent to} a system of interacting particles which can individually deform. To test our predictions, we consider soft spheres as defined in Eq.\ref{pot}, whose
%\be
%\label{num1}
%{\cal U}(\{r\},\{R\})=\frac{1}{2} \sum_{ij} (r_{ij}- (R_i+R_j))^2 \Theta(R_i+R_j- r_{ij}),
%\ee
%where $r_{ij}\!=\!||{\vec r}_i\!-\!{\vec r}_j||$. 
radius follow the internal potential:
\begin{equation}
\label{num2}
\mu\big(\{R\}\big) = \frac{{\bar k}_R}{2} \sum_i  (R_i-R_i^{(0)})^2 \bigg( \frac{R_i^{(0)}}{R_{i}} \bigg)^2,
\end{equation}
where ${\bar k}_R$ is a characteristic stiffness.  We considered a potential diverging as $R_i\!\rightarrow\!0$ to avoid particles shrinking to zero size. To avoid crystallisation we further considered that particles are of two types: for 50\% of them, $R_i^{(0)}\!=\!0.5$ while for the others $R_i^{(0)}\!=\!0.7$. This choice leads to a bimodal distribution of size $\rho(R)$, as shown in Fig.~\ref{f1n}. Our model corresponds to a swap dynamics where swap is allowed only between particles of the same type.   Note that broad mono-modal distributions can be optimised to make swap more efficient while avoiding crystallisation \cite{Ninarello17}, which would be similar to having a very large $\alpha$ in our theoretical description.
The spatial dimension is $d=2$ in our simulations and $k=1$ is our unit stiffness, leading to a simple relation  $\Delta = p$.
 
To study the jamming transition, we consider a pressure-controlled protocol at zero temperature described in the S.I.
The chemical potential of Eq.~(\ref{num2}) must evolve with pressure to maintain a fixed polydispersity. 
As shown in Fig.~\ref{f1n}.B, it can be achieved within great accuracy simply by imposing that $ {\bar k}_R\!=\! p /{\bar \alpha}$, where ${\bar \alpha}$ is a parameter that controls the width $\alpha$ as shown in the inset of Fig.~\ref{f1n}. For this bimodal distribution, $\alpha$ is defined as  $\alpha =(\alpha_1 + \alpha_2)/2$, where $\alpha_1$, $\alpha_2$ are the relative width of each peak in $\rho(R)$. In the limit where the non-swap dynamics is recovered -- which happens when ${\bar \alpha}\rightarrow 0$ --  $\alpha$ and ${\bar \alpha}$ are proportional.

\subsection{Structure and stability}
\begin{figure}[htbp]
\centering
\includegraphics[width=1.0\columnwidth]{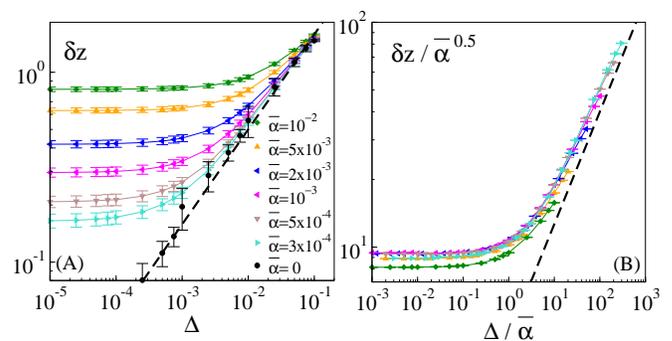}
\caption{(A) log-log plot of $\delta z$ {\it vs.}~$\Delta$ for different values of ${\bar \alpha}$ as indicated in legend. The black dashed line corresponds to $\delta z\!\sim\!\sqrt{\Delta}$.  (B) all these curves collapse if the  y-axis is rescaled by $1/{\bar \alpha}^{0.5}$ and the $x$-axis is rescaled by $1/{\bar \alpha}$. Bars indicate standard deviations.}
 \label{f2n}
\end{figure}

Our central prediction is that for swap dynamics materials must display a larger coordination to enforce stability. %, as sketched in Fig.~\ref{f1}.B. 
This prediction is verified in Fig.~\ref{f2n}.A, which shows $\delta z $ {\it v.s.}~$p$ for various values of ${\bar c}$. Isostaticity is indeed lost and the coordination converges to a plateau as $\Delta$ decreases. Strikingly, we find for the plateau value  $\delta z\!\sim\!\sqrt{\alpha}$,  consistent with a saturation of the stability bound of Eq.~(\ref{10}). This scaling behavior is implied by the scaling collapse in Fig.~\ref{f2n}.B which also confirms that the characteristic overlap  below which swaps affects the dynamics scale as $\Delta\!\sim\!\alpha$. Overall, these results supports that the numerical curves $\delta z(\Delta)$ in Fig.~\ref{f2n}.A  correspond to the marginal stability lines under swap dynamics, shown for different polydispersity (see more on that below).

%summarises our results for the coordination $z$ of the packings. Coordination is defined as $z=2N_c/(N-N_r)$, where $N_c$ is the total number of contacts between the particles and $N_r$ is the number of rattlers (defined as particles with $0$ or $1$ contact). On the right it is shown $\delta z$ {\it vs} $P$ for various values of ${\bar c}$ and on the left it is shown that all these curves collapses to a universal curve when the $\delta z$ is  normalized by $c^{0.5}$ and $P$ by $c$. Numerical results show that configurations generated by our protocol are  marginally stable, which corresponds the equality of Eq.\ref{8}.  First, in the limit  $\alpha/ p\rightarrow 0$, the dynamics without swap is recovered, $\delta z\sim \sqrt{p}$ (shown by the black line in the figure). We emphasize that in our simulations this limit is recovered for ${\bar c} = \infty$, because ${\bar k}_R = \infty$ and the energy cost to modify the radii is infinite.  In the limit where $p<<\alpha$  the marginal stability corresponds to a coordination independent of pressure, as it is confirmed by the plateau in the Figure (\ref{f2n}).

\begin{figure}[htbp]
\centering
\includegraphics[width=1.0\columnwidth]{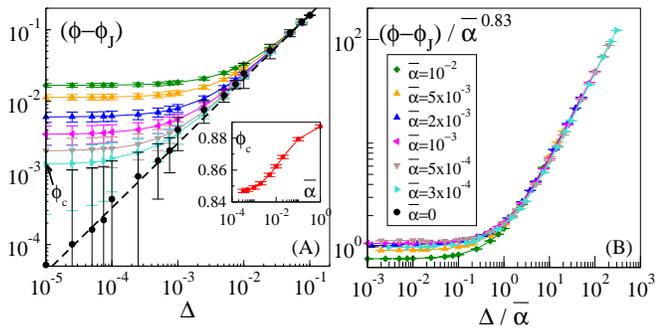}
\caption{ (A) log-log plot of $\phi-\phiJ$ {\it vs} $\Delta$ for different values of ${\bar \alpha}$.  $\phiJ=0.8456$  is defined as the packing fraction for ${\bar \alpha}=0$ and for the smallest pressure we simulate $\Delta=10^{-5}$. Inset: $\phiC$ {\it vs} ${\bar \alpha}$. $\phiC$ is estimated by considering the packing fraction at $\Delta=10^{-5}$. (B) scaling collapse showing that $\phi(\Delta)-\phiJ\approx f(\Delta/{\bar \alpha}) {\bar \alpha}^{0.83}$.}
 \label{f3n}
\end{figure}

\subsection{Packing fraction}
For traditional dynamics, polydispersity tends to have very limited effects on the value of jamming packing fraction  $\phi_J$. We have confirmed this result in the S.I., by showing that although our model can generate very different distributions $\rho(R)$, the values we obtain for $\phi_J$ cannot be distinguished if jamming is investigated using non-swap dynamics.  However, for swap dynamics we expect the situation to change dramatically: since stability requires much more coordinated packings, they presumably need to be denser too.  We denote the jamming packing fraction  for swap $\phi_c\!\equiv\!\lim_{\Delta\rightarrow 0} \phi(\Delta)$. The inset of Fig.~\ref{f3n}.A  confirms that $\phi_c$ increases significantly as $\rho(R)$ broadens. To quantify this effect we consider  $\phi(\Delta,\bar{\alpha})$, as shown in the main panel. Assuming a scaling form for this quantity, and requiring that it satisfies the known results for the jamming transition for  $\Delta\!\gg\!\bar{\alpha}$ implies $\phi(\Delta,\bar{\alpha})\!-\!\phi_J\!=\! f(\Delta/\bar{\alpha})\bar{\alpha}^{\beta}$ where  $f(x)$ is some scaling function and $\beta\!=\!1$. Since the coordination does not change for $\Delta\!\ll\!\bar{\alpha}$, we expect that it is true for the structure overall and for $\phi$, implying that $f(x)\!\sim\! x^0$ as $x\!\rightarrow\!0$. These predictions are essentially confirmed in Fig.~\ref{f3n}.B. Note however that the best scaling collapse is found for $\beta\!=\!0.83\!<\!1$.  These deviations are likely caused by finite size effects, known to be much stronger for $\phi$ than for the coordination or vibrational properties \cite{Ohern03}, and which may thus be present for our systems of $N\!=\!484$ particles.

%
%shows our results for the packing fraction  $\phi=\sum_i \pi R_i^2 /L^2$ of the configurations  at different values of ${\bar c}$ and different pressures. It is represented $\phi$ in terms of the distance to $\phiJ$, which is the Jamming point  for the dynamics without swap. It is defined as the value of $\phi$ for the case where $c = \infty$ and $p=10^{-5}$.  The inset shows the packing fraction at the Jamming point for the dynamics with swap, $\phiC$, as a function of ${\bar c}$. $\phiC$ is defined as the packing fraction at $p=10^{-5}$. We observe that $\phiC$ is higher than $\phiJ$ (value shown by the horizontal dashed line in the inset), evidencing that the swap dynamics is able to generate   denser packing fractions than without the swap. On the right all curves collapses to an universal curve when  $\phi$ is  normalized by $c^{0.83}$ and $p$ by $c$.  If the system size where infinite we should expect the collapse for a $y=\phi . c^1$, right ? The fact that we find the collapse for a smaller value of the exponent, means that we have finite size effects? Is there any other effect?

\begin{figure}[htbp]
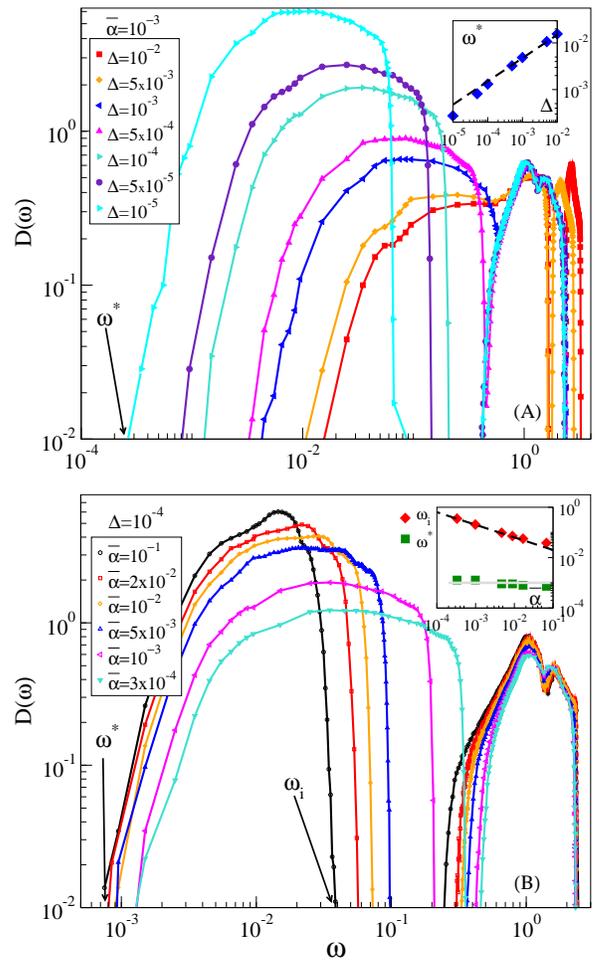

  \centering
  \includegraphics[width=0.89\columnwidth]{gfig4a.eps}
  \includegraphics[width=0.89\columnwidth]{gfig4b.eps}
  \caption{ (A) $D(\omega)$ for different $\Delta$ at fixed  $\bar {\alpha}=10^{-3}$. Inset: $\omega^*$ {\it vs} $\Delta$, where $\omega^*$ is extracted as $D(\omega^*)=10^{-2}$. Dashed line corresponds to the marginality condition $\omega^*\sim \sqrt{\Delta}$. (B)  $D(\omega)$ for different  ${\bar c}$ and fixed $\Delta=10^{-4}$. Inset: $\omega^*$ {\it vs} $\Delta$. Dashed line is the theoretical prediction $\omega_i\sim 1/\sqrt{{\bar \alpha}}$ and the continuous line corresponds to $\omega^*= \sqrt{\Delta}=0.001$.}
  \label{f4n}
\end{figure}

\subsection{Vibrational properties}
We computed the Hessian $H_{\mbox{\scriptsize swap}}$ and diagonalized it (see detailed in SI) to extract the density of states $D(\omega)$,  %The eigenvectors of $H_{\mbox{\scriptsize swap}}$ are the normal modes of the system and the  eigenvalues are the frequencies squared $\omega^2$. The  distribution of these frequencies is the density of states $D(\omega)$.  
 as shown in  Fig.~\ref{f4n}.A  for different pressures at fixed polydispersity. As expected, at low  particle overlap $\Delta$ two bands appear in the spectrum. The lowest-frequency band presents a plateau above some frequency scale  $\omega^*_{\mbox{\scriptsize swap}}$ which satisfies  $\omega^*_{\mbox{\scriptsize swap}}\!\sim\!\sqrt{\Delta}$ as shown in the inset, as expected if the structure were  marginally stable. As shown in S.I.,  in the absence of pre-stress the minimal eigenvalues of the Hessian increase many folds, again a signature of marginal stability \cite{Wyart05a}. Further evidence appears in Fig.~\ref{f4n}.B  showing $D(\omega)$ at fixed $\Delta\!=\!10^{-4}$ for varying polydispersity. $\omega^*_{\mbox{\scriptsize swap}}$ essentially does not depend on $\bar{\alpha}$ as shown in the inset, as expected for marginal packings if the pressure is fixed.  The cut-off frequency $\omega_i$ of the low-frequency plateau scales as   $\omega_i\!\sim\!\sqrt{k_R}\!\sim\!1/\sqrt{\bar{\alpha}}$, as predicted above.

% 
% 
% 
% Fixed ${\bar c}$ means that $\alpha$ is fixed and then Eq.(\ref{9}) predicts $\omega^*_{\mbox{\scriptsize swap}} \sim \sqrt{p}$, which is represented in the inset by the dashed line and is in very good agreement with the numerical points. On the left $D(\omega)$ is shown for different  ${\bar c}$ and fixed $P=10^{-4}$.
%At fixed $P$, $\omega^*_{\mbox{\scriptsize swap}} \sim \sqrt{{\bar c}}$, that is observed in the inset. Interpretation about the  beginning of the plateau and its independence  on ${\bar c}$.
%Each eigenvector of the  Hessian $H_{\mbox{\scriptsize swap}}$  is composed by  the projection of the spatial displacements $\delta \vec{r}_i$ and the fluctuations of the radii $\delta {R}_i$. 

\section{Glass transition}

\begin{figure}[htbp]
  \centering
  \includegraphics[width=1.00\columnwidth]{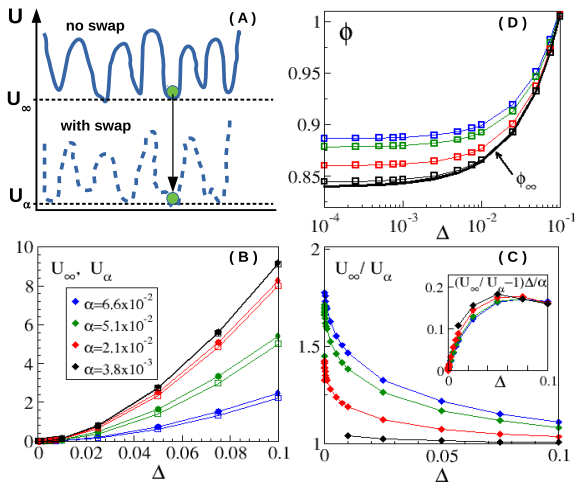}
  \caption{ (a) Sketch illustrating the effect of swap on the energy landscape. Systems rapidly quenched with non-swap dynamics display an energy $U_\infty$. These states
  are unstable to a steepest descent with swap, leading to a lower energy $U_\alpha$: meta-stable states appear at lower energies with swap. (b) Values for $U_\infty$ (full symbols) and  $U_{\alpha}$  (open symbols) as a function of the  dimensionless pressure $\Delta$ imposed during the quenches as a function of $\bar{\alpha}$ as shown in the legend. The ratio $U_\infty/U_{\alpha}$  is shown in (c), it is stronger near jamming but remains significant even far for jamming is the system is sufficiently poly-disperse. \maa{As shown in the inset, in relative terms the shift of the energy of inherent  structures $(U_\infty-U_{\alpha})/U_{\alpha}$ is proportional to $\alpha$ and inversely proportional to $\Delta$ when $\Delta$ is large enough.} (d) The packing fraction $\phi_c$ obtained after swap is turned on is larger than $\phi_\infty$ obtained for non-swap dynamics, an effect that is stronger near jamming.  }
  \label{ske}
\end{figure}

We now turn to the glass transition, which always takes place at a sizeable distance form the jamming transition \cite{Ikeda12}: for example for hard discs,
$\phi_g\approx 0.78$ and $\phi_c\approx 0.85$. A similar difference of packing fraction occurs by compressing soft spheres at overlap $\Delta\approx 0.05$, as illustrated in Fig.\ref{ske}(d).
From the arguments above, we expect that if the poly-dispersity is sufficiently large, vibrational properties will be strongly affected even far away from jamming, in particular near the glass transition. 

The direct consequence of this fact is that the energy landscape will be affected by swap, which will in turn affect the glass transition. At high energy configurations are unstable --- they are saddles with many unstable directions --- whereas below some characteristic energy minima appear. However, since stability is strictly more demanding with swap, this characteristic energy must be reduced when swap is allowed for. We prove this point in Fig.\ref{ske}.(a,b), where inherent structures of energy $U_\infty$ are obtained after using a steepest descent for non-swap dynamics. These configurations are not stable for our generalised steepest descent that let particles deform, which leads to configurations of energy  $U_{\alpha}\!<\! U_\infty$. This effect is stronger near jamming in relative terms as shown in Fig.~\ref{ske}.(c), but remains significant away from jamming if the poly-dispersity is broad enough. It corresponds for example to a reduction of energy of 25\% for $\Delta=0.05$ for our $\alpha=0.06$. \maa{ We show in the inset of that panel that the relative shift of energy induced by swap $(U_\infty-U_{\alpha})/U_{\alpha}$ is proportional to $\alpha$ and inversely proportional to $\Delta$ when $\Delta$ is large enough, in   consistence with what we found for the structure in Eq.(\ref{8}). }

\begin{figure}[htbp]
  \centering
  \includegraphics[width=0.60\columnwidth]{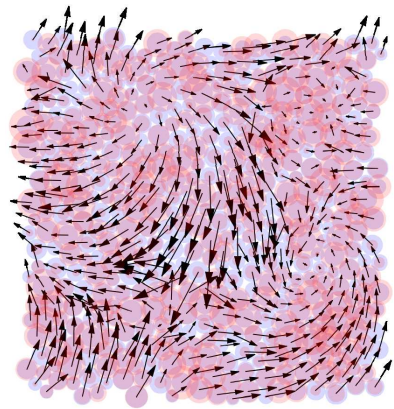}
  \caption{Example of soft mode with $\omega\!\approx\!0.078$ for  $\bar{\alpha}=1$ and $\Delta\!=\!10^{-2}$. Blue discs  indicate the initial particles radii,  red discs the new radii induced by motion along that mode. Arrows represent the displacements $\delta \vec r_i$  of the particles  multiplied by 4 for visualisation.}
  \label{f5n}
\end{figure}

Thus as the temperature is lowered in these liquids, the Goldstein temperature where activation sets in will be smaller when swap is allowed for. 
This analysis thus predicts an entire range of temperature where the non-swap dynamics is slowed down by activation, whereas the swap dynamics can flow along unstable modes. 
\maa{More quantitatively, we predict the shift of glass transition temperature $\Delta T_g/T_g$ induced by swap to be proportional to $\alpha$, in consistence with the observation that very broad distributions  lead to large swap effects \cite{Ninarello17}. We also  predict that $\Delta T_g/T_g$ is inversely proportional to the distance to jamming $\Delta$
when this quantity is well-defined (e.g. for soft spheres, but also to some extent  for Lennard-Jones potentials \cite{Wyart05b,Xu07}) and large enough. }
In real space, the unstable modes that render activation useless involve both translational degrees of freedom as well as swelling and shrinking of the particles.  We show an example of such a mode in Fig.~\ref{f5n}, corresponding to the softest mode of the generalised Hessian we obtain with parameters   $\alpha=0.06$  and $\Delta\!=\!10^{-2}$. It illustrates that the particle displacements are not necessarily divergent free when swap is allowed, since the system can locally compress or expand by changing the particle sizes.  

This interpretation of swap acceleration is consistent with the observation that the dynamics is less collective with swap at the temperature where the non-swap dynamics is activated, since the system can rearrange locally without jumping over barriers if there are enough unstable modes. Collective dynamics is expected only when these modes become less abundant at lower temperatures. Likewise, we expect the Debye-waller factor to be larger with swap, since the vibrational spectrum is softer. Note that these arguments are not restricted to finite range interactions. We expect them to apply as well to Lennard-Jones potentials for example, where the abundance of degrees of freedom {\it vs.}~the number of strong interactions is also known to affect the vibrational spectrum \cite{Wyart05b,Xu07}.

\section{Hard sphere systems} 

Our arguments above consider the energy landscape. For interactions potentials which are very sharp, non-linearities induced by thermal fluctuations are important,
and the vibrational properties of a glassy configurations at finite temperature $T$  can differ significantly from those of its inherent structure obtained by quenching it rapidly. Here we consider the extreme case of hard spheres where the energy is always zero, and  cannot be used to define vibrational modes.  Instead, by averaging on vibrational time scales within a glassy configuration, a local free energy can be defined \cite{Brito06,Brito09,Altieri16} where particles that are colliding within that state interact with a logarithmic potential. This description is exact near jamming and systematic deviations are expected away from it \cite{Altieri18}. However in practice, the Hessian defined from this free energy captures  well the fluctuations of particle positions and the vibrational dynamics throughout the glass phase \cite{Brito09}.  (This procedure can be pursued to include thermal effects in soft spheres as well \cite{DeGiuli15}).

\begin{figure}[htbp]
\centering                              
\includegraphics[width=1.0\columnwidth]{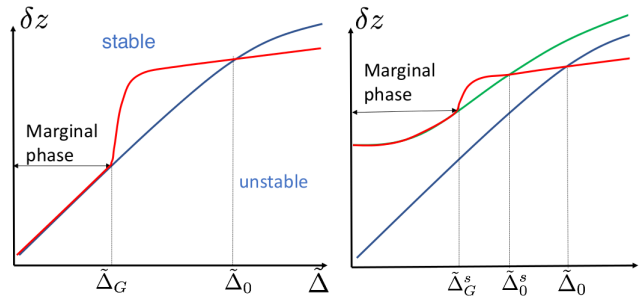}
\caption{Log-log representation of the stability diagram in the coordination $\delta z\!=\! z -  z_c$ and ${\tilde \Delta}$ plane   for continuously poly-disperse  thermal hard spheres with non-swap (Left) and swap (Right) dynamics. Note that for hard spheres ${\tilde \Delta}$ is independent of temperature and vanishes at jamming. In the Left panel, the blue line separates mechanically stable and unstable configurations. The red line indicates the trajectory of a system under a slow compression. When ${\tilde \Delta}$ decreases toward the onset gap ${\tilde \Delta}_0$, metastable states appear and the dynamic becomes activated and spatially correlated. In the glass phase, the red trajectory will depend on the compression rate, but will eventually reach the blue line at some (rate-dependent) ${\tilde \Delta}_G$. When this occurs, a buckling or Gardner transition takes place where the material becomes marginally stable, leading to a power-law relation between ${\tilde \Delta}$ and $\delta z$. Right: for swap dynamics, stability is more demanding and is achieved only on the green line, which differs strictly from the blue one.  Thus the onset gap decreases to some value ${\tilde \Delta}_0^{s}\!<{\tilde \Delta}_0$: the dynamics become activated and correlated at larger densities, shifting the position of the glass transition.  Marginality is still expected beyond some pressure ${\tilde \Delta}_G^s$, but leads to  plateau value for the coordination, indicating that isostaticity is lost.}
  \label{f1}
\end{figure}

Stability and vibrational properties can be computed in terms of this Hessian for non-swap dynamics \cite{Brito06,Brito09}. Salient results are shown in the simplified diagram of Fig \ref{f1}. Once again, two key determinant of stability  are the typical gap between interacting particles ${\tilde \Delta}\equiv T/(p R_0^{d})$ relative to the particle radius,  and  the excess coordination $\delta z$, where the coordination is defined from the network of particles that are colliding within a glassy state.   A marginal stability line  separates stable and unstable configurations, as illustrated in Fig.\ref{f1}, whose asymptotic behaviour follows $\delta z\!\sim\! {\tilde \Delta}^{(2+2\theta)/(6+2\theta)}$ where $\theta\!\approx\!0.41$ \cite{DeGiuli15}. (Strictly speaking, this line will depend slightly on the system preparation, but this dependence is expected to be modest, and is irrelevant for the present discussion).  

Under a slow compression the system follows a line (in red) in the $({\tilde \Delta},\delta z)$ plane. Mechanical stability is reached only for some ${\tilde \Delta}\!<\! {\tilde \Delta}_0$, a characteristic onset gap where the dynamic crosses-over to an activated regime where vibrational modes become stable, in consistence with Goldstein's proposal. In these materials, deeper in the glass phase the system eventually returns to the stability line, and undergoes  a sequence of buckling events that leave it  marginally stable \cite{Wyart05a,Brito09,DeGiuli14b}. Marginal stability  implies the presence of soft elastic modes (that differs from Goldstone modes) up to nearly zero frequency, and fixes the scaling properties of both  structure and  vibrations  as jamming is approached \cite{Wyart05a,DeGiuli14b}. These results, valid in finite dimensions, have been quantitatively confirmed in infinite dimension calculations \cite{Charbonneau14,Franz15,Altieri16}. In that case, the point where buckling sets in was argued to be a sharp transition, coined Gardner, where the free energy landscape fractures in a hierarchical way \cite{Charbonneau14}, as supported by numerical studies  \cite{Berthier16}. For very rapid quenches, it was argued that the entire glass phase should be marginal \cite{Brito09,Charbonneau14}.

How is this picture affected by swap? Our arguments for the generalised Hessian of the soft sphere system  essentially go through unchanged for the generalized Hessian of the free energy in the hard sphere system. Once again, stability becomes more demanding with swap, and the marginal stability line is shifted to higher coordination in the $({\tilde \Delta}, \delta z)$ plane as represented in the right panel of Fig. \ref{f1}.  Thus the glass transition is shifted toward higher packing fractions. At smaller gap ${\tilde \Delta}$ (corresponding to the approach of jamming), stability implies  $\delta z\!\geq\!\alpha^{(2+2\theta)/(6+2\theta)}$, again implying that isostaticity is lost with swap. We conjecture that, just as for soft particles, marginal stability is reached in the glass phase, which would correspond to a Gardner transition in infinite dimensions.

\section{Conclusion}  In swap algorithms, the dynamics is governed by an effective potential  ${\cal V}(\{r\},\{R\})$ that describes both the particles interaction and  their ability to deform. As a result, we have shown that vibrational and elastic properties are  softened when swaps are allowed for, while thermodynamic quantities are strictly preserved (when thermal equilibrium is reached). This result supports that the cross-over temperature  $T_0$ where  mechanical stability appears and dynamics becomes activated must be reduced with swap with $T_0^{\mbox{\scriptsize swap}}\!<\! T_0$, leading to a natural explanation as to why the glass transition occurs then at a lower temperature  $T_g^{\mbox{\scriptsize swap}}\!<\! T_g$. %Our analysis thus supports that near the glass transition, vibrational properties control  the nature of elementary rearrangements and their associated activation barriers, instead of thermodynamics quantities such as growing static lengths. 
Secondly, swap must strongly affect the structure of the glass phase. This is particularly striking near the jamming transition that occurs in hard and soft spheres, where we predict that well-known key properties such as isostaticity must disappear. We have confirmed these predictions numerically, and found that for rapid quenches the effective potential  ${\cal V}(\{r\},\{R\})$ appears to be marginally stable throughout the glass phase.

Concerning the glass transition,
our work does not specify the mechanism by which activation occurs in glasses, but it does  support that swap delays the temperature where activation is required to relax, which potentially explains several  previous observations of swap algorithms \cite{Berthier16b,Gutierrez15,Ninarello17,Berthier17}.  Possible theories to describe the mechanism by which activation occurs in glasses include elastic \cite{Dyre06} and facilitation models \cite{Garrahan02}. We do think however that theories based on a growing thermodynamic length will be hard to reconcile with the notion that some collective modes do not see any barriers at all.

Our analysis also makes additional qualitative testable predictions. By increasing continuously the width $\alpha$ of the radii distribution $\rho(R)$, we predict that $T_g^{\mbox{\scriptsize swap}}(\alpha)$ will smoothly decrease, while $T_g(\alpha)$ should be essentially unchanged\maa{, with  $(T_g(\alpha)-T_g^{\mbox{\scriptsize swap}}(\alpha))/T_g(\alpha)\propto \alpha$ and more specifically $\propto \alpha/\Delta$ for soft spheres}. Furthermore,   many studies have analyzed correlations between dynamics and vibrational modes, see e.g.~\cite{Broderix00,Grigera02,Widmer-Cooper08,Brito09},  which can be repeated to relate the swap dynamics to the spectrum of the effective potential  ${\cal V}(\{r\},\{R\})$. Near $T_g$, we predict the latter to have more abundant modes at low or negative frequencies than the much studied Hessian of the potential energy, and its softest modes  to be better predictors of further relaxation processes. Lastly, the present analysis suggests that adding additional degrees of freedom (such as changing the shape of the particles, and not only their size) will increase even further the difference between swap and non-swap dynamics.

 %\maa{Explain how to add a Langevin noise to the dynamics of $\{r\}$ and  $\{R\}$ to reach equilibrium}

 Finally, we have shown that ultra-stable glasses can be built on the computer, simply by {\it descending} along the effective potential ${\cal V}(\{r\},\{R\})$.  As illustrated in Fig.\ref{f1}, these configurations must sit strictly inside the stable region of the regular dynamics (i.e. at a finite distance from the blue line). As a consequence, the usual potential energy landscape ${\cal U}(\{r\})$ around the obtained configurations does not display excess soft anomalous modes at very low frequency,  even near the jamming transition: these modes are gapped. This result must hold for the ground state too (which must be stable toward swap) and by continuity also for low-temperature equilibrated states. It may explain why marginal stability (and the Gardner transition leading to it) could  not be observed in protocols where  a thermal quench was used 
from swap-generated configurations \cite{Scalliet17}. It would be very interesting to see if other well-known excitations of low-temperature glassy solids are also gapped in these configurations, including  two-level-systems, reported to be almost absent in experimental ultra-stable glasses \cite{Queen13}.

\begin{acknowledgements}
We thank  L.~Berthier,  G.~Biroli, M.~Cates, M. Ediger and F.~Zamponi for discussions. E.~L.~acknowledges support from the Netherlands Organisation for Scientific Research (NWO) (Vidi grant no.~680-47-554/3259). M.~W.~thanks the Swiss National Science Foundation for support under Grant No. 200021-165509 and the Simons Foundation Grant ($\#$454953 Matthieu Wyart). 
\end{acknowledgements}

%\bibliography{../bib/Wyartbibnew}
\bibliography{Wyartbibnew}

\appendix
\section{SUPPLEMENTAL MATERIAL}

This supplemental material (SM) provides: $(i)$ descriptions of the numerical model, protocols and methods used to generate athermal packings under swap dynamics at different pressures, $(ii)$ a computation of the Hessian of the potential energy, together with explanations about how pre-stress affects the vibrational modes, and $(iii)$ a discussion about the effect of the radii distribution generated by swap dynamics on the value of the packing fraction of our athermal packings obtained while freezing the degrees of freedom associated with particles' radii.

\subsection{Numerical model, protocols and methods}

We employ systems of $N\!=\!484$ particles in a square box in two dimensions. The total potential energy depends upon the particles coordinates $\{r\}$ and radii $\{R\}$, as
\begin{equation}\label{foo00}
{\cal V}(\{r\},\{R\}) = {\cal U}(\{r\},\{R\}) + \mu(\{R\})\,.
\end{equation}
The pairwise potential term reads
\begin{equation}\label{foo03}
{\cal U}(\{r\},\{R\})=\frac{k}{2} \sum_{ij} \big(r_{ij}- (R_i+R_j)\big)^2 \Theta(R_i+R_j- r_{ij}),
\end{equation}
where $k$ is a stiffness, set to unity, $r_{ij}$ is the distance between the $i\th$ and $j\th$ particles, and $\Theta(x)$ is the Heaviside step function. The chemical potential associated with the radii is
\begin{equation}\label{foo01}
\mu(\{R\})\!=\!\frac{{\bar k}_R}{2}\!\sum_i\! \big( R_i\!-\! R_i^{(0)}\big)^2 \bigg( \frac{R_i^{(0)}}{R_{i}} \bigg)^2 \!\equiv\! \sum_i \mu(R_i,R_i^{(0)}\!)
\end{equation}
where $\bar{k}_R$ is the stiffness of the potential associated with the radii $\{R\}$, that serves as a parameter in our study, and is set as described below. $R_i^{(0)}$ denotes the intrinsic radius of the $i\th$ particle. In each configuration we randomly assigned $R_i^{(0)}\!=\!0.5$ for half of the particles, and {$R_i^{(0)}\!=\!0.7$} for the other half. The mass $m$ of particles, and that associated with their fluctuating radii, are all set to unity. Vibrational frequencies should be understood as expressed in terms of $\sqrt{k/m}$, and pressures in terms of $k$. 

Configurations in mechanical equilibrium at zero temperature and at a desired target pressure $p_0$ were generated as follows; we start by initializing systems with random particle positions at packing fraction $\phi=1.2$, and set the initial radii to be $R_i\!=\! R_i^{(0)}$. We then minimize the total potential energy ${\cal V}(\{r\},\{R\})$ at a target dimensionless  pressure $\Delta_0\!=\!10^{-1}$ using a combination of the FIRE algorithm \cite{Bitzek06} and the Berendsen barostat \cite{Berendsen84}, see futher discussion about the latter below. Each packing is then used as the initial conditions for sequentially generating lower pressure packings, as demonstrated in Fig.~\ref{sm1}. Following this protocol, we generated 1000 independent packings at each target  {dimensionless} pressure, that ranges from  $\Delta_0\!=\!10^{-1}$ up to $\Delta_0\!=\!10^{-5}$. For each target pressure, we set the stiffness $\bar{k}_R$ of the chemical potential of the radii according to {$\bar{k}_R\!=\! p_0/\bar{\alpha}$, and vary $\bar{\alpha}$ systematically between $3\times 10^{-4}$ and $1$. During minimizations we calculate a characteristic \emph{net} force scale $F_{\mbox{\tiny typ}}\!\equiv\!\big(\sum_{i}||\vec{F}_{i}||^2/N\big)^{1/2}$, where $\vec{F}_i\!=\!-\partial {\cal V}/\partial \vec{r}_i$ is the net force acting on the $i\th$ particle, whose coordinates are denoted by $\vec{r}_i$. A packing is considered to be in mechanical equilibrium when $F_{\mbox{\tiny typ}}$ drops below $10^{-8}\Delta_0$. %{\color{red}{IF THE PRESSURE DROPS WAY BELOW THE TARGET PRESSURE, THE SYSTEM IS NOT A SOLID... DOES THIS HAPPEN AT ALL? ARE THOSE CONFIGURATIONS DISCARDED OF?}} Fig.(\ref{sm1}) shows the pressure of the system as a function of the number of iterations for one particular initial condition to exemplify the protocol.
%It gives an idea of how fast the pressure converges to the 

\begin{figure}[htbp]
  \centering
  \includegraphics[width=0.90\columnwidth]{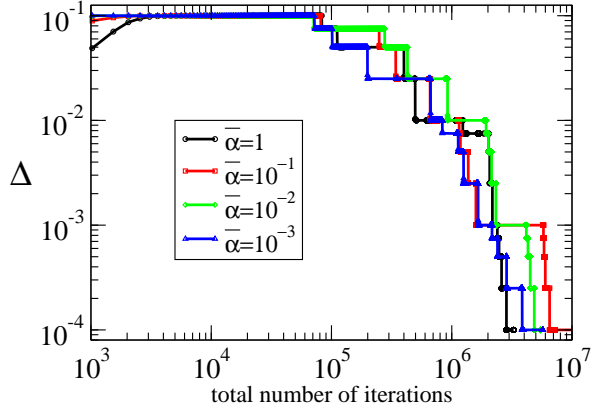}
  \caption{Dimensionless pressure $\Delta$ as a function iteration number for a packing-generating simulation starting from one particular initial condition. Each step of the staircase shape of the signal corresponds to the production of a packing at some desired target pressure. The criterion for convergence to mechanical equilibrium at each pressure is explained in the text. We produced packings ranging from $\Delta\!=\!10^{-1}$ to $\Delta\!=\!10^{-5}$.}
  \label{sm1}
\end{figure}

\begin{figure}[htbp]
  \centering
    \includegraphics[width=0.90\columnwidth]{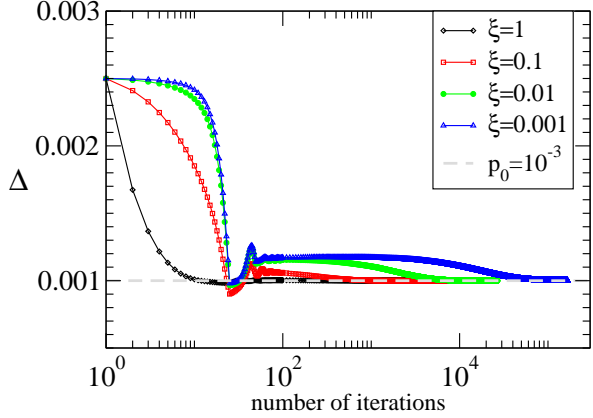}
  \caption{Instantaneous dimensionless pressure $\Delta$ as a function of the number of iterations for one initial condition, taken to be $\Delta\!=\!0.025$. Here we set ${\bar \alpha}\!=\!10^{-3}$, fix the target pressure to $\Delta_0\!=\!10^{-3}$, and observe the form of the convergence of the instantaneous dimensionless pressure to the target  dimensionless pressure, for different values of the Berendsen barostat parameter $\xi$ (see definition in text).}
  \label{sm2}
\end{figure}

{\bf Berendsen barostat parameter:~} The FIRE algorithm \cite{Bitzek06} features equations of motion which are to be integrated as in conventional MD simulations. We exploit this feature and embed the Berendsen barostat~\cite{Berendsen84} in our Verlet integration scheme \cite{Allen89}. This amounts to scaling the simulation cell volume by a factor $\chi$, calculated as
\begin{equation}
  %\chi=1- \xi \delta t(p_0-p)\,,
  \chi=1- \xi \delta t(\Delta_0-\Delta)\,,
\end{equation}
where $\delta t$ is the (dynamical) integration time step, and $\xi$ is a parameter that determines how quickly the instantaneous dimensionless pressure converges to the target dimensionless pressure~\cite{Allen89}. In Fig.~\ref{sm2} shows the $\xi$-dependence of the convergence of the instantaneous dimensionless pressure $\Delta $ to the target value $\Delta_0$. Below $\xi=0.01$, the behavior of $\Delta$ as a function of iteration number is similar. We therefore set $\xi=0.01$ througthout this work.

\begin{figure}[htbp]
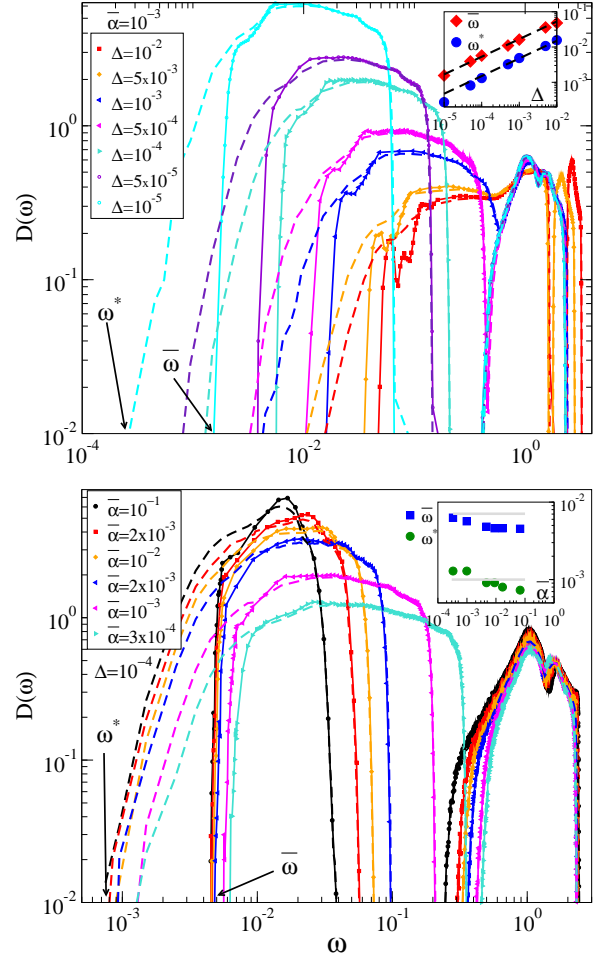

  \centering
  \includegraphics[width=0.89\columnwidth]{fig3a_sm.eps}
  \includegraphics[width=0.89\columnwidth]{fig3b_sm.eps}
  \caption{Top: Density of states $D(\omega)$ for different dimensionless pressures $\Delta$ at fixed ${\bar \alpha}\!=\! 10^{-3}$, with the pre-stress term included (dashed lines) and excluded (continuous lines). Inset: characteristic frequencies $\omega^*$ and $\bar{\omega}$ (as marked in the main panel) {\it vs} $\Delta$. Dashed line correspond to the marginality condition $\omega^*\!\sim\!\sqrt{\Delta}$ and $\bar{\omega}\!\sim\!\sqrt{\Delta}$. Bottom: $D(\omega)$ for different  ${\bar {\alpha}}$ and fixed $\Delta=10^{-4}$. Note that for these curves $\omega^*{}^2/\bar{\omega}^2\approx 5\%$, indicating that the system is very close to marginal stability. }
  \label{smDw}
\end{figure}

\subsection{Computation of $H_{\mbox{\scriptsize swap}}$}
The total potential  energy ${\cal V}(\{r\},\{R\})$ of our model system is spelled out in Eqs.~(\ref{foo00})-(\ref{foo01}). We next work out the expansion of ${\cal V}$ in term of small displacements $\delta \vec{r}_i$ of particle positions, and small fluctuations $\delta R_i$ of the radii, about a mechanical equilibrium configuration with energy ${\cal V}_0$, as
\begin{eqnarray}\label{foo02}
\delta {\cal V} \equiv {\cal V} - {\cal V}_0 & \simeq &\  \  \sFrac{1}{2}\sum_{ij}\delta\vec{r}_i\!\cdot H_{ij}\!\cdot\!\delta\vec{r}_j  \nonumber \\
& & + \sFrac{1}{2}\sum_{ij}\delta R_iQ_{ij}\delta R_j \nonumber \\
& & + \sum_{ij}\delta R_iT_{ij}\cdot\delta\vec{r}_j \,,
\end{eqnarray}
where $H_{ij}\!\equiv\!\partial^2{\cal V}/\partial\vec{r}_i\partial\vec{r}_j$, $Q\!\equiv\!\partial^2{\cal V}/\partial R_i\partial R_j$, and $T_{ij}\!\equiv\!\partial^2{\cal V}/\partial\vec{r}_i\partial R_j$. The expansion given by Eq.~(\ref{foo02}) can be written using bra-ket notation as
\be
\delta  {\cal V} = \sFrac{1}{2}  \bra{\delta \ell}H_{\mbox{\scriptsize swap}}\ket{\delta\ell},
\ee
where  $\ket{\delta\ell}$ is a $(d\!+\!1)N$-dimensional vector which concatenates the spatial displacements $\delta \vec{r}_i$ and the fluctuations of the radii $\delta {R}_i$: $(\delta \vec{r}_1, \delta \vec{r}_2, .., \delta \vec{r}_N, \delta {R}_1,...~\delta {R}_N)$. The operator $H_{\mbox{\scriptsize swap}}$ can be written as:
\[
H_{\mbox{\scriptsize swap}} = \left[
\begin{array}{cc}\Bigg[\hbox{\makebox[10ex]{$H_{\mbox{\tiny$Nd,\! Nd$}}$}} \Bigg]&\Bigg[\hbox{\makebox[5ex]{$T_{\mbox{\tiny$Nd,\! N$}}$}}\Bigg] \\ \big[\hbox{\makebox[10ex]{$T^T_{\mbox{\tiny$N,\! Nd$}}$}}\big]&\big[\hbox{\makebox[5ex]{$Q_{\mbox{\tiny$N,\! N$}}$}}\big]\end{array}\right]\,.
\]
The elements of the submatrix $H_{{\scriptscriptstyle\rm Nd,Nd}}$ can be written as tensors of rank $d\!=\!2$ as
\begin{eqnarray*}
H_{ij}=\delta_{\langle ij \rangle} \bigg(\frac{ k(r_{ij}\!-\! R_i\!-\! R_j)}{2r_{ij}}{\vec n}_{ij}^{\perp}\otimes{\vec n}_{ij}^{\perp} + \frac{k}{2} {\vec n}_{ij}\otimes{\vec n}_{ij}\bigg) \nonumber \\
 + \delta_{i,j} \sum_{l}\bigg(\frac{ k(r_{il}\!-\! R_i\!-\! R_l)}{2r_{il}}{\vec n}_{il}^{\perp}\otimes{\vec n}_{il}^{\perp} + \frac{k}{2} {\vec n}_{il}\otimes{\vec n}_{il}\bigg),
\end{eqnarray*}
where $\vec{n}_{ij}$ is a unit vector connecting between the $i\th$ and $j\th$ particles, $\vec{n}_{ij}^\perp$ is a unit vector perpendicular to $\vec{n}_{ij}$, $\otimes$ is the outer product, $\delta_{\langle ij \rangle}\!=\!1$ when particles $i,j$ are in contact, $\delta_{i,j}$ is the Kronecker delta, and the sum is taken over all particles $l$ in contact with particle $i$. The elements of the submatrix $Q_{\mbox{\tiny$N,\! N$}}$ are scalars given by:
\begin{equation}
  Q_{ij}=\delta_{\langle ij \rangle}k +  \delta_{i,j}\bigg(\sum_{\langle l\rangle} k + \frac{\partial^2 \mu(R_i,R_i^{(0)})}{\partial R_i^2}\bigg)\,.
\end{equation}
The matrix $T_{\mbox{\tiny$N,Nd$}}$ is not diagonal and each element can be expressed as a vector with two components given by:
\begin{equation}
  T_{ij}= -\delta_{\langle ij \rangle}k{\vec n}_{ij} -  \delta_{i,j} \sum_{l} k{\vec n}_{il}.
\end{equation}
The eigenvectors of $H_{\mbox{\scriptsize swap}}$ are the normal modes of the system, and the eigenvalues are the vibrational frequencies squared $\omega^2$. The distribution of these frequencies is known as the density of states $D(\omega)$.

%%\begin{eqnarray}
%% \delta  {\cal V} \approx \sum_{\langle ij \rangle} k(r_{ij}-R_i-R_j)\frac{[(\delta \vec{r}_j-\delta \vec{r}_i).{\vec n}_{ij}^{\perp}]^2}{2r_{ij}^{eq}} + \nonumber \\ \sum_{\langle ij \rangle} \frac{k}{2} [(\delta \vec{r}_j-\delta \vec{r}_i).{\vec n}_{ij}]^2.
%%\end{eqnarray}
  
%\subsection{C. Density of states with and without swap}
%\begin{figure}[htbp]
%  \centering
%  \includegraphics[width=0.99\columnwidth]{graf-c1000_swap_2d_separa.eps}
%  \caption{comparison}
%  \label{sm2}
%\end{figure}

%%\subsection{D. Finite sizes effects: packing fraction ???}

\begin{figure}[!ht]
  \centering
  \includegraphics[width=0.89\columnwidth]{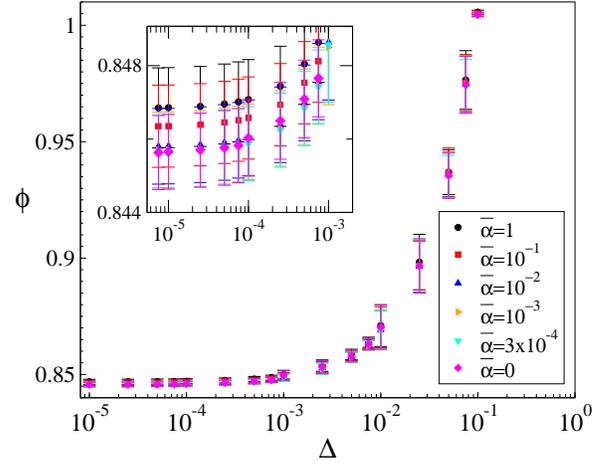}
  \caption{Packing fraction $\phi$ as function of dimensionless pressure $\Delta$ measured for packings in which radii are not allowed to fluctuate, and whose distribution of radii is borrowed from swap packings generated at $\Delta\!=\!10^{-4}$ and at different values of ${\bar \alpha}$, as indicated by the legend. The inset shows a zoom into very small dimensionless pressures, demonstrating that the value of $\phi_c$ depends very weakly on the borrowed distribution of radii of the swap packings, as determined by the parameter ${\bar \alpha}$.}
  \label{smPhi}
\end{figure}

{\bf Effect of the pre-stress on the vibrational modes:~}
When a system of purely repulsive particles is at mechanical equilibrium, forces $f_{ij}$ are exerted between particles in contact. These forces give rise to a term in the expansion of the energy, of the form
\begin{equation}
-\frac{1}{2}\sum_{\langle ij \rangle} \frac{f_{ij}}{r_{ij}}\big((\delta\vec{r}_j - \delta\vec{r}_i)\cdot\vec{n}_{ij}^\perp\big)^2\,,
\end{equation}
often referred to as the ``pre-stress term''. For plane waves, it can be shown that the energy contributed by this term is very small. However, for the soft modes present when the system is close to the marginal stability limit, it can be shown that this term reduces the energy of the modes by a quantity proportional to the pressure \cite{Wyart05a}. Marginal stability corresponds to a buckling transition where the destabilising effect of pre-stress exactly compensate the stabilising effect of being over-constrained. In this scenario where two effects compensate each other, the eigenvalue of the softest (non-Goldstone) modes of  the Hessian  in the absence of pre-stress $\bar{\omega}^2$ must be much larger than $\omega^*{}^2$ computed when pre-stress is present. 
To demonstrate this, we have calculated the density of states for systems while including and excluding the pre-stress term. The results are shown in Fig.~\ref{smDw}, where it is found that near jamming $\omega^*{}^2/\bar{\omega}^2\approx 5\%$, which is consistent with what previously found for the traditional jamming transition \cite{DeGiuli14b} and supports that the system is very close to (but not exactly at) marginal stability.

%\vspace{0.1cm}

\subsection{Packing fraction}

In the main text we have shown that the jamming packing fraction $\phi_c$ generated using the swap dynamics increases when $\rho(R)$ broadens, i.e. for smaller values of the parameter $\bar{\alpha}$ that controls the stiffness of the potential energy associated with the radii. Here we compare the dependence of the packing fraction on pressure as measured for systems in which the radii are not allowed to fluctuate. In addition, in this test we borrow the distribution of radii $\rho(R)$ from swap-packings generated at $p\!=\!10^{-4}$, and at various values of the parameter $\bar{\alpha}$, varied between 1 to $\infty$ (the latter corresponds to disallowing particle radii fluctuations). Packings were generated using the total potential energy as given by Eq.~\ref{foo03} (with the radii $R_i$ considered to be fixed), and using the same protocol and numerical methods used to generate the swap packings. The results are shown in Fig.~\ref{smPhi}, where it can be seen that the value of $\phi_c$ is essentially the same for any borrowed $\rho(R)$ from the swap packings.

\end{document}